\newif\ifpdf
\ifx\pdfoutput\undefined
\pdffalse 
\else
\pdfoutput=1 
\pdftrue \fi

\documentclass{elsart}
\usepackage{amsmath}
\usepackage{amssymb}
\usepackage{verbatim}
\IfFileExists{myowntimes.sty}
	{\usepackage{myowntimes}}{\usepackage{times}\usepackage{mathrsfs}}

\newcommand{\Deltac}{\Delta_{\text{\rm c}}}
\newcommand{\Thetac}{\varTheta_{\text{\rm c}\,}}

\begin{document}
\begin{frontmatter}
\title{\fontsize{16}{22}\selectfont%
Comment on: ``Theory of the evaporation/condensation\\transition of equilibrium droplets in finite volumes''}
\author[LosAngeles]{Marek~Biskup,$^\dagger$}
\author[LosAngeles]{Lincoln Chayes} and
\author[Prague]{Roman Koteck\'y}
\address[LosAngeles]{Department of Mathematics, UCLA, Los Angeles, California, USA}
\address[Prague]{Center for Theoretical Study, Charles University, Prague, Czech Republic}
\begin{abstract}
We examine some aspects of the recent results by K.~Binder~\cite{Binder}.
The equilibrium formation/dissolution of droplets in finite systems is discussed in the context of the canonical and the grand canonical distributions.
\end{abstract}

\end{frontmatter}

\renewcommand{\thefootnote}{}
\footnotetext{\hglue-4mm\fontsize{10}{12}\selectfont$^\dagger$ corresponding author; address: 
UCLA Mathematics Department, Box 951555, Los Angeles, CA 90095-1555; email: \textsf{\fontsize{8.5}{9}\selectfont biskup@math.ucla.edu}}
\footnotetext{\hglue-4mm\fontsize{10}{12}\selectfont\copyright\ \ Copyright rests with the authors. Reproduction
of the entire article for non-commercial purposes is permitted without charge.}
\renewcommand{\thefootnote}{\arabic{footnote}}

\noindent
In the last few years, a considerable number of computer experiments, for instance \cite{Kosterlitz,Chinese,MS,PH,PS}, carefully performed on systems exhibiting phase coexistence have underscored the need for a better understanding of the droplet formation/dissolution phenomena. 
In this context, some early analyses~\cite{Binder-Kalos,Sethna,DS} pointed to the existence of a volume dependent (mesoscopic) scale at which droplets first appear. 
(Specifically, it was argued that in a system of volume~$L^d$, one does not observe droplets below the linear scale of~$L^{d/(d+1)}$.)
Recently, a detailed quantitative description of the actual droplet formation/dissolution in \emph{closed} equilibrium systems has been accomplished~\cite{Neuhaus-Hager,BCK1}.
For instance, the following was shown in~\cite{BCK1} regarding a gas-liquid system in volume~$L^d$ and the number of particles fixed to a value exceeding that of the ambient gas by amount~$\delta N$:
\begin{enumerate}
\item[(1)]
There is a dimensionless parameter~$\Delta$ proportional to $(\delta N)^{(d+1)/d}/L^d$ and a critical value $\Deltac=\Deltac(d)$, such that no droplet forms for $\Delta<\Deltac$, while there is a \emph{single} droplet of liquid phase when $\Delta>\Deltac$.
\item[(2)]
The fraction $\lambda_\Delta\in[0,1]$ of the excess particles subsumed by the droplet depends on~$\Delta$ via a \emph{universal} equation which depends only on dimension (and which is otherwise independent of the details of the system).
\item[(3)]
The minimal fraction $\lambda_{\Deltac}=2/(d+1)$ is stricly positive, so when the droplet first forms, it indeed has volume of the order~$L^{d^2/(d+1)}$.
\end{enumerate}
Further investigations permitted a rigorous proof of the above conclusions in the context of the two-dimensional Ising lattice gas at all temperatures below critical~\cite{BCK2} (as well as a rigorous derivation of the Gibbs-Thomson formula under certain conditions~\cite{BCK3}).

The intriguing circumstances concerning the systems with coexisting phases were the subject of a recent paper by K.~Binder~\cite{Binder} wherein the existence of the mesoscopic scale for droplet formation/dissolution was re-derived by phenomenological arguments.
Two additional conclusions of interest were reached~in~\cite{Binder}:
\begin{enumerate}
\item[(4)]
A signature discontinuity in the \emph{intensive} variable relative to the setup at hand, that is, the \emph{magnetic field} in a spin system and the chemical potential in a liquid/gas system.
\item[(5)]
The scaling window for the ``rounding'' of this discontinuity in finite systems.
\end{enumerate}
While we are somewhat uneasy about the derivation of~(5)---which in our opinion poorly accounts for the possible influence of lower order corrections---we will focus our attention on conclusion~(4).
The substance of this conclusion is apparently novel and warrants further investigations, particularly because of the purported connection with other ``unconventional'' phase transitions, see reference~35 of~\cite{Binder}.
We will concentrate on the Ising ferromagnet in a~$d$-dimensional volume~$L^d$. 
Although the magnetic language is used in~\cite{Binder}, the lattice-gas interpretation is invoked to label the ensembles: 
The constrained ensemble with fixed total spin (i.e., fixed magnetization) will be referred to as the ``canonical'' ensemble, whereas the ``grand canonical'' ensemble will denote the usual distribution 
in which the magnetization is allowed to fluctuate.

Inherently in its nature, the magnetic field is a quantity associated with (and \emph{adjustable} only in the context of) the ``grand canonical'' ensemble.
This leads us to our first question:
How does the purported discontinuity reflect itself in the ``grand canonical'' ensemble?
To address this issue, let us investigate the problem of the Ising magnet in a box of linear dimension~$L$, at the temperature $T<T_{\text{\rm c}}$, external field~$h$ and plus boundary conditions. 
The cases of interest are $h\le0$ with $|h|\ll1$, which are the only conditions under which the system might nucleate a droplet. 
Denoting by~$R$ the linear scale of the purported droplet, the magnetic gain from its formation would be of the order of~$hR^d$, while the surface cost would scale as~$R^{d-1}$. 
Obviously, the two costs balance out for $R\sim1/|h|$, so if $L\gtrsim R$ permits~$R$ to exceed a constant times~$1/|h|$, such a droplet \emph{will} form and otherwise it won't.
This, of course, is exactly the basis for classical nucleation theory.

Notwithstanding any doubts as to the validity of the above reasoning, the preceding setup has been the subject matter of some rigorous analysis, see~\cite{SS1,SS2,GS1,GS2}. In particular, the following two-dimensional result was established in~\cite{SS2}: Consider the setup as described (with plus boundary conditions and $h<0$), with $|h|\to0$ and $L\to\infty$ in such a way that~$|h|L$ tends to a definite limit, denoted by~$B$. Then there is a $B_0>0$ (which can be calculated in terms of system characteristics), such that the following holds:
\begin{enumerate}
\item[$\bullet$]
If $B<B_0$, there are no droplets and the entire box is in the plus phase.
\item[$\bullet$]
If $B>B_0$, a large droplet of minus phase fills most of the box leaving only a small fraction of the plus phase in the corners. 
\end{enumerate}
Thus, whenever the droplet forms, it subsumes the \emph{bulk} of the system.
Similar (albeit weaker) theorems were proved in~\cite{GS1,GS2} for all~$d>2$.

These results are of direct relevance and lead to the following inescapable conclusion: In the context of the ``grand canonical'' distribution, there is no window of opportunity for the formation of a \emph{mesoscopic} droplet. Explicitly, whenever conditions permit the existence of a ``droplet'' in the system, it occurs on the \emph{macroscopic} scale. Ostensibly, one might still hope for the occurrence of some signature event when the magnetic field lies in (or in the vicinity of) Binder's gap. However, this is not the case: Binder has calculated the edges of the forbidden region,
$$
H_t^{(1)}=(d+1)\frac{m_{\text{coex}}(T)}{\chi_{\text{coex}}(T)}V_d^{-\frac{d-1}{d+1}}\Bigl(\frac{S_d}{2d}\Bigr)^{\frac d{d+1}}\Bigl(\frac L{\xi_{\text{coex}}(T)}\Bigr)^{-\frac d{d+1}}c^{\frac d{d+1}}
$$
and
$$
H_t^{(2)}=(d-1)\frac{m_{\text{coex}}(T)}{\chi_{\text{coex}}(T)}V_d^{-\frac{d-1}{d+1}}\Bigl(\frac{S_d}{2d}\Bigr)^{\frac d{d+1}}\Bigl(\frac L{\xi_{\text{coex}}(T)}\Bigr)^{-\frac d{d+1}}c^{\frac d{d+1}},
$$
where~$V_d$,~$S_d$ and~$d$ are geometrical constants,~$m_{\text{coex}}(T)$,~$\chi_{\text{coex}}(T)$ and~$\xi_{\text{coex}}(T)$ is notation for the magnetization, susceptibility and the correlation length, respectively, and $c=f_s(T)\chi_{\text{coex}}(T)m_{\text{coex}}(T)^{-2}\xi_{\text{coex}}(T)^{-1}$---with~$f_s(T)$ denoting the surface tension---is a dimensionless ratio (canceling out the superfluous~$\xi_{\text{coex}}(T)$'s!) which presumably tends to a constant as $T\to T_{\text{c}}$. But, at the end of the day, both edges satisfy $H_t^{(i)}\sim L^{-d/(d+1)}$, which, we emphasize, is \emph{deep} inside the droplet dominated regime.

On the basis of the above deposition, it appears that conclusion~(4) has absolutely no bearing on finite-volume systems described by the ``grand canonical'' ensemble. 
The question is then: How to interpret the magnetic field and its purported discontinuity otherwise? 
As is clear from the outset, some non-standard interpretation will be necessary since the only physical framework in which the phenomenon occurs is the ``canonical'' ensemble. 
In the context of the Ising model in volume~$L^d$ and plus boundary conditions, the latter describes the constrained distribution where the overall magnetization~$M_L$ is restricted to a \emph{single} value. 
(Here, as goes without saying, the external field~$h$ in the Hamiltonian simply drops out of the problem.)
To achieve a droplet of minus phase, there has to be a \emph{deficit} in the magnetization away from the preferred value of~$M_L$. 
In such circumstances, the general results discussed in the introductory paragraph imply the existence of a sharp constant~$\Thetac$ (related to~$\Deltac$) such that no droplet will be created for deficits less than $\Thetac L^{d^2/(d+1)}$, while, for deficits larger than $\Thetac L^{d^2/(d+1)}$, a non-trivial fraction of the deficit will condense into~a~droplet.

Let us now attempt to elucidate how a magnetic field could have arisen in the derivations of~\cite{Binder}.
Of course, in the ``canonical'' ensemble, we are always entitled to calculate the (finite-volume) free energy as a function of the magnetization. 
As is necessarily implied by the nature of the above droplet formation/dissolution phenomenon, this function has two branches depending on what type of configurations bring the decisive contributions:
\begin{enumerate}
\item[$\bullet$]
For magnetizations with a deficit less than~$\Thetac L^{d^2/(d+1)}$, configurations with no mesoscopic droplets.
\item[$\bullet$]
For magnetizations with a deficit in excess of~$\Thetac L^{d^2/(d+1)}$, configurations with a single appropriately-sized droplet.
\end{enumerate} 
It is not much of a surprise that a \emph{cusp} will form at the point where the two branches come together. It appears that the values~$H_t^{(1)}$ and~$H_t^{(2)}$, which are enunciated explicitly in~\cite{Binder}, are just the one-sided derivatives of the free energy---with respect to magnetization---at this cusp.

Unfortunately, the physical significance attributed to the values~$H_t^{(1)}$,~$H_t^{(2)}$, their difference and their ratio in~\cite{Binder} is perhaps a bit overplayed. 
Indeed, following the dogma of \emph{bulk} thermodynamics, the~``$H$'' is proclaimed to be the natural canonical conjugate of the magnetization and, as such, it is deemed to be the appropriate measure of the response of the system to the change of the magnetization.
However, here we deal with a system exhibiting mesoscopic phenomena and, more importantly, inhomogeneities.
In such systems the meaning of a conjugate variable is rather murky because the standard interpretations of the thermodynamic potentials are only clear in the thermodynamic limit, under the auspices of the equivalence of ensembles.
Consequently, for the system at hand, the \emph{primary} response functions should be the~``$H$'s'' associated with the parts of the system outside and inside the droplet, which we note are perfectly analytic functions of the corresponding magnetizations.
On the basis of the latter response functions, and the knowledge of the droplet size, the overall~``$H$'' considered in conclusion~(4) can immediately be reconstructed.
But, even if this quantity could be conveniently accessible numerically, its actual meaning is at best secondary.

We would like to remark that, in our opinion, the probabilistic language of large-deviation theory provides some additional and worthwhile perspectives in these situations. In the terminology of large-deviation theory, the actual free energy can be conveniently expressed as an infimum of a simple function over what seems to be the natural parameter here: The \emph{fraction of the deficit absorbed by the droplet}. With this parametrization, the relevant calculations of~\cite{Binder}, including the jump in the derivative at the formation point, fit on the back of the proverbial envelope. We refer to~\cite{BCK1,BCK2,Neuhaus-Hager} for more details but we do not wish to overstate our case.

\section*{Conclusions}
\noindent
The conclusion/moral is self-evident. 
In general, given a function, we are always entitled to take its Legendre transform and express it in terms of the conjugate variables. 
In the context of equilibrium statistical mechanics, these transforms are invaluable because the equivalence of ensembles allows for the uninhibited two-way flow of information. 
For instance, if a particle system is studied at a fixed density then, except at points of phase transitions, we know everything about the fluctuating ensemble with the chemical potential adjusted to produce this density. 
Even more interesting---and even more useful---are the points of thermodynamic discontinuities. 
If one ensemble has a forbidden gap (say the particle density in the grand canonical distribution), then forcing the ``parameter value'' into the gap is essentially guaranteed to have interesting consequences in the other ensemble (e.g., phase separation). 

But, the equivalence of ensembles is a mathematical---not to mention phy\-si\-cal---fact only in the \emph{thermodynamic limit}. 
In finite volume, as the droplet formation/dis\-solu\-tion phenomenon dramatically illustrates, the various ensembles are \emph{not} equivalent. 
In these contexts, the assignment of physical---not to mention mathe\-ma\-tical---significance to the conjugate variables is of dubious value. 
We suspect that this is the generic situation when ``phase transitions'' on a mesoscopic scale are the object of study. 
We believe that the dramatic inequivalence of ensembles in finite volume is the \emph{signature} of interesting phenomena taking place below the macroscopic scale.

It is worth pointing out that, in the present context, the natural thermodynamic quantity which exhibits the signature jump is the good old \emph{energy density}.
There are several advantages to the use of this quantity as opposed to e.g. the magnetic field considered in~\cite{Binder}.
To list a couple, first, there is no numerical difficulty in the dynamical construction of the energy histogram and, second, there is no theoretical dispute in the interpretation of this quantity.
Some previous efforts to exhibit the behavior of the energy density can be found in \cite{PS,PH,Neuhaus-Hager}; but, here we emphasize that the actual energy should be measured directly.
Notwithstanding, if the physics of interest concerns \emph{droplets}, it appears most natural to look for the droplet itself.
This is evidently numerically feasible~\cite{Sethna,PH} and, presumably, permits the exhibition of \emph{all} the secondary commodities.



\begin{thebibliography}{100}



\bibitem{Binder}
K.~Binder,
\textit{Theory of evaporation/condensation transition of equilibrium droplets in finite volumes}, Physica~A \textbf{319} (2003) 99-114.

\bibitem{BCK1}
M.~Biskup, L.~Chayes and R.~Koteck\'y, 
\textit{On the formation/dissolution of equilibrium droplets}, 
Europhys. Lett. \textbf{60:1} (2002) 21--27.

\bibitem{Binder-Kalos}
K.~Binder and M.H.~Kalos, 
\textit{Critical clusters in a supersaturated vapor: Theory and Monte Carlo simulation}, 
J.~Statist. Phys. \textbf{22} (1980) 363--396.

\bibitem{BCK2}
M.~Biskup, L.~Chayes and R.~Koteck\'y, 
\textit{Critical region for droplet formation in the two-dimensional Ising model},
submitted (\textsf{\fontsize{9}{9}\selectfont http://xxx.lanl.gov/abs/math.PR/0212300})

\bibitem{BCK3}
M.~Biskup, L.~Chayes and R.~Koteck\'y, 
\textit{A proof of the Gibbs-Thomson formula in the droplet formation regime}, 
submitted (\textsf{\fontsize{9}{9}\selectfont http://xxx.lanl.gov/abs/math-ph/0302031})


\bibitem{DS}
R.L.~Dobrushin and S.B.~Shlosman, 
\textit{Large and moderate deviations in the Ising model},
In:~\textit{Probability contributions to statistical mechanics}, 
pp.~91--219, Adv. Soviet Math., vol.~20, Amer. Math. Soc., Providence, RI, 1994.

\bibitem{GS1}
P.E.~Greenwood and J.~Sun, 
\textit{Equivalences of the large deviation principle for Gibbs measures and critical balance in the Ising model},
J.~Statist. Phys. \textbf{86} (1997),  no.~1-2, 149--164.


\bibitem{GS2}
P.E.~Greenwood and J.~Sun, 
\textit{On criticality for competing influences of boundary and external field in the Ising model},
J.~Statist. Phys. \textbf{92} (1998), no.~1-2, 35--45.


\bibitem{Kosterlitz}
J.~Lee and J.M.~Kosterlitz, \textit{Finite-size scaling and Monte
Carlo simulations of first-order phase transitions}, \textrm{Phys.
Rev. B} \textbf{43} (1990) 3265--3277.

\bibitem{Chinese}
J.~Machta, Y.S.~Choi, A.~Lucke, T.~Schweizer and L.M.~Chayes,
\textit{Invaded cluster algorithm for Potts models}, \textrm{Phys.
Rev. E} \textbf{54} (1996) 1332--1345.

\bibitem{MS}
T.~M\"uller and W.~Selke, \textit{Stability and diffusion of
surface clusters}, \textrm{Eur. Phys.~J.~B} \textbf{10} (1999)
549--553.


\bibitem{Neuhaus-Hager}
T.~Neuhaus and J.S.~Hager, \textit{$2d$ crystal shapes, droplet 
condensation and supercritical slowing down in simulations of 
first order phase transitions}, cond-mat/0201324.

\bibitem{PH}
M.~Pleimling and A.~H\"uller, 
\textit{Crossing the coexistence line at constant magnetization}, 
\textrm{J.~Statist. Phys.} \textbf{104} (2001) 971--989.

\bibitem{PS}
M.~Pleimling and W.~Selke, 
\textit{Droplets in the coexistence region of the two-dimensional Ising model}, 
\textrm{J.~Phys.~A: Math. Gen.} \textbf{33} (2000) L199--L202.

\bibitem{Sethna}
B.~Krishnamachari, J.~McLean, B.~Cooper and J.~Sethna,
\textit{Gibbs-Thomson formula for small island sizes: Corrections for high vapor densities}, 
\textrm{Phys. Rev.~B} \textbf{54} (1996) 8899--8907.

\bibitem{SS1}
R.H.~Schonmann and S.B.~Shlosman,
\textit{Complete analyticity for $2$D Ising completed},
Commun. Math. Phys. \textbf{170} (1995), no. 2, 453--482.

\bibitem{SS2}
R.H.~Schonmann and S.B.~Shlosman,
\textit{Constrained variational problem with applications to the Ising model},
J.~Statist. Phys. \textbf{83}  (1996),  no. 5-6, 867--905.


\end{thebibliography}
\end{document}